\documentstyle[epsfig,12pt]{article}
\begin{document}                
\begin{center}
{\Large\bf Cosmic GRB energy-redshift relation and Primordial flares as possible
 energy source for the central engine}
\end{center}
\begin{center}
{\Large\bf K. M. Hiremath}
\end{center}
\begin{center}
{\em Indian Institute of Astrophysics, Bangalore-560034, India,
E-mail : hiremath@iiap.ernet.in}
\end{center}

\begin{abstract}
By considering similar observed properties of  gamma ray
bursts (GRB) and solar flares
 with the prevailing physical conditions in the cosmic environment, the
following study suggests that most likely and promising energy source 
for the central engine which triggers GRB may be due to primordial flares,
solar flare like phenomena, at the sites of inter galactic or inter galactic
clusters in the early universe. The derived energy-redshift relation ,
$E = E_{0}{(1+z)}^3$ (where $E$ is the amount of energy released,  $z$
is the redshift of GRB and $E_{0}$ is a constant which is estimated
to be $\sim \,10^{52}$ ergs ), from the simple flare mechanism,
is confirmed from the least square fit with the observed energy-redshift
relation. Some of the physical parameters like length scale, strength of
magnetic field, {\em etc.,} of the flaring region of the GRB are 
estimated.
\end{abstract}

\section{Introduction}

 Since the discovery of GRB, the source of the
central engine which triggers this phenomenon remains elusive till today. 
To date, there are more than hundred 
theoretical models on GRB ( {\em} see the bibliography web site, 
{\em http://ssl.berkeley.edu/ipn3/index.html}, compiled by Hurley). 
Based on the insufficient data, in the initial period 
of observations, many models on GRB favored energy source of galactic origin.
However, recent observations are not consistent with these models 
and the models which explain that source of GRB could be of 
extragalactic origin 
are most favorable (Piran 1999).  Though, these models can explain some of the 
observed phenomena, other outstanding observed properties such as energy
source of the central engine and isotropic distribution of the GRB in the 
universe remain to  be explained. 

Presently there are two main schools of thoughts on the
central engine of the GRB : (i) neutron star-neutron
star ( or neutron star -black hole) binary mergers
and, (ii) hypernovae. Though very few of the observations
appear to favor these models, they have the following difficulties
(McLaughlin {\em et al.} 2002).
 (a) The small spatial
offsets between GRB counterparts \& their host galaxies
(Bloom, Kulkarni \& Djorgovski 2002) are
hard to reconcile with the merging neutron stars
which supposed to have large offsets from the center
of the host galaxy. (b) Though there appears to be a strong
association between the GRBs and the star formation rates,
collapser (hypernovae) model has the following difficulties.
An emission line has been found (Antonelli, et al. 2000; 
Piro et al. 2000) in some x-ray afterglows of the gamma 
ray bursts (GRB) whose energy is roughly consistent with Fe K$\alpha$
at the redshift of the host. This is clearly against
the hypernovae model, since hypernovae (supernovae)
primarily produce nickel (Woosley \& Weaver 1995), not iron. 
Even if one believes that hypernovae models explain 
the source of central engine, it is not clear to us
 that such a large-scale phenomena accrete material onto a black 
hole and form GRB events on the time scales of few seconds
to minutes.
Hence, some other physical phenomenon
may be responsible for the central engine which triggers 
 GRB events.
It is found from this investigation that, primordial flares (in the regions of
inter galactic or inter galactic clusters), {\it i.e.,} solar flare like
phenomena in the early history of the universe, may be most promising energy source for the creation of GRB.
Though similar studies (Qu \& Wang 1977; Stecker and Frost 1973;
Vahia and Rao 1988) indicate relevance of stellar flares as the energy
source of GRB, the  observed spectra of GRB imply that
GRB must be of extragalactic origin and hence it is unlikely that stellar flares
may be energy source for the cosmic GRB.

Previous studies (Stecker \& Frost 1973; Vahia and Rao 1988)
show many similar observed properties between the solar
flare like phenomenon and the GRB event like phenomenon.
We add following additional similarities of GRB events with the solar
flares : (i) the source size is compact,
(ii) time scales is $\sim$ seconds to minutes, (iii) radiation is observed
in most of the electromagnetic spectrum, (iv) non-thermal energy spectrum
and (v) polarization $< 30 \% $.
 It is also interesting to note that
emission line around 5 kev in GRB000214 is almost
similar to the emission line at 5 kev in solar
x-ray flare (Zirin 1988)
which along with the ideas presented in our paper
give much credence that GRB events may be solar
flare like phenomenon in the cosmic environment
on larger scales.  From the observed similar properties
of the neighborhood solar flares and the distant cosmic GRBs
and, the following energy-redshift relation derived from the physics
of flare phenomenon, we propose in the present study
that the central engine which triggers cosmic GRB transient events
may be solar flare like phenomenon on larger scale in the
early history of the universe. During that epoch the universe was very 
active and the first stars and effectively galaxies were either begin to form 
or might have formed (Larson 1999). 

Presently it is believed (Priest 1981; Haisch \& Strong 1991; Parker 1994) 
that the source of energy produced in the solar flares is due to phenomenon 
called {\em magnetic reconnection} in a very compact region wherein oppositely 
directed magnetic 
flux, in the limit of finite electric conductivity, annihilate each other
and releasing required amount of flare energy with the acceleration
of highly energetic particles (Lenters \& Miller 1998; Tsuneta and Naito 1998).
In the following we give a brief introduction of theory of magnetic
reconnection invoked for the study of the solar flares.

\section{  Magnetic reconnection and Primordial Flares}
Let oppositely directed magnetic flux of large length scale $L$ 
are merged with  inflow velocity $v_{in}$. This merging of flux will form a current
sheath of thickness $\delta$. Then the law of magnetic induction according
to MHD description is 
$$
{{\partial B } \over{\partial t}} = curl ( v \times B) + \eta {\nabla}^2 B \, ,  \eqno (1)
$$
where $B$ is strength of the magnetic field, $v$ is the velocity and $\eta$ 
is the magnetic diffusivity of the plasma. The first term in the above equation
is due to convective flow and second term represents the magnetic diffusion
of the plasma. Outside the region of reconnection, magnetic diffusivity
is very low and magnetic field is glued to the plasma and moves
along with the plasma. That is magnetic field is frozen to
the plasma. This condition of infinite electric conductivity fails in
the region of magnetic field reconnection by producing very high
gradients of currents and electric fields. Dissipation of these
strong currents leads to annihilation of magnetic field 
in the region of magnetic reconnection where  
 steady state exists so that convective and resistive
terms in equation (1) are equal. The amount of energy released by 
the annihilation of magnetic field $B$ and cube of length $L$
is estimated to be $\sim$ $L^{3}B^{2}$. The ratio of convective to
resistive term called "magnetic Reynolds number" is given as follows
$$
R_{m} = {v_{in} \delta \over { \eta}}       \, , \eqno (2)  
$$
where $\delta$ is the thickness of reconnection region. The assumption of 
incompressibility and conservation of flux yields
$$
v_{out} = {L \over {\delta}} v_{in} = v_a    \, , \eqno (3)
$$
where $v_{out}$ is out flow velocity, $v_{in}$ is the inflow
velocity  and,  $v_a$ is the ambient Alfven 
wave  velocity whose perturbations are
perpendicular to the field line and travel along it. 

Outside the region of magnetic reconnection, the convective term
in equation (1) dominates over the resistive term. That is $R_{m}>>1$ and
hence the electric field is non-dissipative. However, inside the layer of 
thickness $\delta$, $R_{m}<<1$ and
, the electric field $E = \eta J$ is dissipative which leads to 
domination of kinetic effects.  For example, two dimensional simulations 
(Brown 1999) indicate that magnetic flux and 
electron flow decouple in the accelerating region resulting in 
acceleration of electron beams  by ejecting with super-Alfvenic velocity.
\subsection{Primordial Flares}
Primordial flares means that flares of extragalactic
origin when the universe was in the nascent stage
with high magnetic activity like the
sun's atmosphere. We expect the flares to occur in the optically 
thin medium situated either in the regions of inter galactic
or inter galactic clusters. The length scales are of cosmic 
dimension and, flares of extragalactic origin are produced by the 
oppositely directed magnetic 
flux due to peculiar motion of  cosmic bodies such as galaxies or cluster of 
galaxies or near the sites of supernova ejections from the galaxy.
The region of magnetic reconnection is assumed to be formed
by merging of oppositely directed magnetic field which is
of primordial origin. The typical observed peculiar velocity fields
of galaxies are $\sim 10^7$ cm/sec and length scales of magnetic elements $\sim 10^{24} $ cms (Giovaneli 1998).
It is expected that in order to reproduce the observed magnetic fields
 $\sim \mu$ G in galaxies, galaxy clusters and inter galactic clusters 
(Kroneberg 1994; Vallee {\em et al.} 1987; Lesch and Chiba 1997; Bagchi 1988;
 Taylor and Perley 1993), a primordial field of $\sim 10^{-9} $ G is required.
\smallskip

 Based on the simple flare model described
in \S  2 and available observed informations on GRB, in the following, first
we derive the GRB energy-redshift relation and compare with the
observed energy-redshift values. From the linear least square fit
to the observed data, we estimate the average energy liberated
from the GRB events.  we also estimate size, strength of the magnetic field 
and thickness of  the reconnection layer
at the site of GRB formation.  Using this information and conservation
of mass and flux, we estimate strength of the magnetic field of the region 
at the present epoch ($z = 0$). Then we compare estimated strength
of magnetic field and velocity of outflow $v_{out}$ from the site
of flare region with the strength of the 
observed large-scale magnetic field and peculiar velocity of the galaxies.

\section {GRB Energy-Redshift Relation}

We know that the amount of energy released by annihilation of magnetic
field of strength $B$ and cube of length scale $L$ is $E \sim L^3 B^2$. 
We assume that in the early universe, on the average, large-scale
currents are zero. This implies that  currents $J = curl{\bf B} = 0$, 
where $\bf B$ is the magnetic field.
This relation leads to the potential fields 
and hence a dipole field like structure is the likely dominant mode
which still may be existing in the cosmic environment.
Dipole field like structure varies as $\sim L^{-3}$, where
L is comoving length at the time of emission.
Apply the redshift correction,
then comoving length $L$ is related to length scale
$L_{0}$ at the present epoch as $L_{0} = L(1+z)$ . Next plug in this
relation in the energy $E$ ($\sim L^3 B^2$) and from the 
conservation of magnetic flux, we
get the following relation
$$
E = E_0 (1 + z)^3 .  \eqno (4)
$$
Here  $z$ is the redshift of the GRB events and $E_{0}$ is constant of 
energy to be determined
from the observed data. This relation implies that {\em the energetics
of the GRB events is approximately directly proportional to the cube of 
their redshift values}. This means the GRB events appear to be more energetic
in the early universe compared to the universe which has small $z$ values.

 In order to check validity of thus derived GRB
energy-redshift relation, we fit a general relation $E = E_{0}(1+z)^{M}$, 
where $E_{0}$ (ergs) and $M$ are constants to be determined from the 
least square fit of the observed energy and redshift data. 
By linearizing the general form , we obtain the relation
$Y = C + MX$, where $Y=log(E)$, $X=log(z+1)$, $C=log(E_{0})$.
We choose 12 GRB events which have known energy and redshift 
values available from the web site  http://www.aip.de/\~jcg/grbrsh.html. 
After subjecting to the least square fit to the observed data, 
 we found the constant coefficients to be $C = 51.6$
which implies $E_{0}=10^{51.6}$ ergs, $M=3.05$ and significance of the 
$\chi^{2}$ value is $ > 99 \%$ . 
Basically $\chi^{2}$ ( $= \sum_{i} [(N_{i}-n_{i})^2/n_{i}]$, where $N_{i}$
is the number of events observed in ith bin, and $n_{i}$ is the
number of expected events) value
tests how our computed  energy values agree with the observed energy
distribution of GRBs. It is to be noted that in order to compute
the $\chi^{2}$, uncertainties in the data is not necessary. 
 Since uncertainties are
presently not available in the observationally determined
redshift values, we could not determine the uncertainties
in the determined parameters $C$ and $M$ respectively.
On the other hand, if sufficient data is available, one
can determine the uncertainties from the property of the standard deviations
 determined from the observed data (Press {\em et al.} 1992).
 In Fig 1., we illustrate the 
GRB energy-redshift relationship.
The observed energy values are denoted by signs of star and 
continuous line is a linear least square fit to the data.
It is crucial to be noted that the coefficient, $M=3.05$, determined from the 
fit of the empirical data is very close to coefficient $M=3.0$,  
according to our expectation that {\em primordial flares
may be the energy sources of the central engine} which trigger
GRB events.

 Owing to a very small data sample
of the observed log(energy)-log(z+1) data, we obtained
a nominal but significant value of the correlation coefficient ($\sim 50\%$ )
in our analysis. In fact one may argue that finding a luminosity that rises 
with distance shouldn't be surprising
for a flux limited sample. However, from the following
analysis using large sample of Fenimore \& Ramirez-Ruiz (astroph-
0004176) luminosity data, such a statistically significant log(energy)-log(z+1)
relationship exists. In any case, one may argue that, the redshift uncertainties 
are usually either extremely
small, or the redshifts in the literature are a lower limit on the true
redshift (for some absorption redshifts). In order to avoid such doubts
from the reader's mind, we enhanced the reliability of the statistics
in the following way.

We consider a large sample of the redshift-energy data from the time
variability of the GRBs obtained from Fenimore \& Ramirez-Ruiz (astroph-
0004176). For our analysis, we chose such a sample
(z values in the steps of 0.1)
of 42 log(energy)-log(z+1) values ( Fenimore and Ramirez-Ruiz
 astro-ph0004176) and fitted with the
generalized law $E = E_{0}(1+z)^{M}$. More data enabled
us to estimate the uncertainties in the following 
parameters : $E_{0} = 10^{50.6\pm 0.27}$,  $M=3.01 \pm 0.31$, 
coefficient of correlation  $90 \%$ and significance of $\chi^{2}$
 value is $>99\%$. Though the values of $E_{0}$ determined
from both the fits appear to be different, in near future, as we collect
more observations of energy-redshift data, both the
values of $E_{0}$ come closer. In Fig 2., we
illustrate the inferred energy-redshift relationship.
Note that value of the index $M$ determined from
the observed energy-redshift data and inferred
energy-redshift data is same. This extra analysis of
least-square fitting shows that our proposed conjecture
( "the energy-red shift relation obtained from the simple
flare mechanism is similar to the observed energy-redshift
relation") is right.

\section{Estimation of Physical parameters of the GRB}
The observations show that the time scales of GRB phenomena vary
from $10^{-3}$ sec to $10^{3}$ sec. For the known redshift of GRBs,
 we find that average time scale is $\sim$  100 sec.
By constraining that out flow velocity in the current sheet should
not exceed the velocity of light and in order to produce average energy $E_{0}$ 
$\sim 10^{52}$ ergs, the length $L$ of the cube should be $\sim 10^{12}$ cms
 and field strength in the reconnection region to be $\sim 10^{8} $ G. 
It is crucial to be noted that the estimated field strength is 
close to the strength of magnetic field inferred from
the observations (Piran 1999) of GRB spectrum in absorption lines.  
\smallskip
If we assume $ L = \delta $, then from equation (3), this implies 
that $v_{in} = v_{out} = v_{a} $. 
Thus from the constrained outflow velocity ($10^{10}$ cm/sec) which is 
also equal to Alfven velocity,  $v_{a}$, and from inferred strength of the 
magnetic field, density of the reconnection region is found to be 
$\sim 10^{-5} \, g/cm^{-3}$. 

Let us now reverse back from the phase of the early universe, 
using inferred length scale and strength of 
magnetic field in region of reconnection, and ask what could be the strength of
 magnetic field at the present epoch ( $z = 0$) that might have been 
distributed over large length scales. If so, does it match with strength of 
the observed large-scale magnetic fields in the universe. Conservation of
flux leads to a relation 
$$
B_{0} = B_{present} {L^{2} \over R_{0}^{2}}   \, , \eqno (5) 
$$
where $B_{0}$ and $R_{0}$ are strength and length scale of the magnetic field 
at the present epoch ($z = 0$), $B_{present}$ and L are strength 
and length scale of magnetic field in the
reconnection region in the earlier epochs. If we take the observed large-scale
magnetic field which have typical length scales 
$R_{0} \sim 10 \, kpc \sim 10^{22} cm$,
we get $B_{0} \sim 10^{-12}$ G which is almost similar
to the expected field strength of primordial origin at the present epoch.
Hence the typical rate of GRB is expected to be $\sim R_{0}/v_{in}
 \sim 10^{5}$ years which is very close to the observed
rate of GRB $\sim 10^6$ years. 

\section{ Conclusions and discussion }
Our conclusions of this study are as follows :
\smallskip

 By considering the similar observed properties of GRB and
solar flares, it is proposed that the energy source of the 
 gamma ray bursts of extragalactic origin is similar to
the energy source which triggers solar and stellar transient activity phenomena
like flares.  From the simple theory of flare mechanism, we derived the 
energy-redshift
relation and compared with the distribution of observed energy and
redshift values. From the linear least square fit, with a high significance of
$\chi^2$ value, it is found that the derived flare
energy-redshift relation very well matches with the
observed energy and redshift relation confirming that {\em most likely and 
promising energy source for the central engine which triggers GRB phenomena
may be due to primordial flares, similar to solar flare like phenomena,
which might have occurred in the early universe.} From the least
square fit, we estimated the average energy $E_{0}$ of
the GRB events to be $\sim \, 10^{52}$ 
ergs and inferred the physical parameters of the flaring region.
The inferred results indicate that the source
region requires length scale of $\sim 10^{12}\, cms$, strength   
of the annihilating magnetic field $\sim 10^{8}$ G and the outflow
velocity of the plasma may be $\sim 10^{10} cm/sec$ if we
assume that thickness is equal to length scale of the reconnection
region. By the theory of conventional flare mechanism and conservation
of mass, we also estimated density of the reconnection region to be
$\sim 10^{-5} \, gram\,cm^{-3}$. Finally, by taking the observed typical
length scale of the cosmic environment at the present epoch ($z = 0$)
and conservation of magnetic  
flux, we estimated the strength of the primordial magnetic field
which is almost similar to expected strength of the magnetic field
of primordial origin. Lastly, we estimated typical rate of GRB
per galaxy to be $\sim 10^{5} $ yrs.

\smallskip

In this study, we inferred from flare theory that in order to get
the required observed GRB energy, length scale of source region may be 
$\sim 10^{12}$ cms and strength of the magnetic field $\sim 10^{8} $ G. 
The difficulty here is how to annihilate such a large scale region within few
seconds. The answer lies in the instabilities (Hood 1986) created by 
attaining such a structure of the magnetic fields. Once instability starts, 
reconnection starts with in few seconds, accelerate the particles very close
to  velocity of light and shock structures would  be formed which
follows the creation of non-thermal spectrum. 

One may get the following
doubts : (i) the model no longer explains why essentially all
well-localized GRBs are found in the stellar field galaxies, (ii) the 
model doesn't explain any of the apparent evidence for links between actively
star-forming regions and GRBs and (iii) the model doesn't explain any
of the rich phenomenology of long-lasting GRB afterglows. For all these
questions, we have following answers. We can not say that essentially all
well-localized GRBs are found in the stellar field galaxies. As explained
in the introduction, in fact, there are small offsets between GRB counter
parts and their host galaxies. Moreover, recent analysis (Schaefer, 1999;
Band,  Hartmann and Shaefer 1999; Mirabal, {\em et al.,} 2002) disfavor GRB events associated with the star-forming regions where
one would expect optically thick medium. That means the optically
thin environment like the sun's atmosphere is essential for the 
GRB production. Infact the studies (Schaefer, 1999; Band,  Hartmann and Shaefer 1999) suggest one of the possibilities that GRBs might reside
in the inter galactic space which we also propose as the probable site
for the GRB origin. As for the second doubt, it is only apparent
evidence and not all the observed GRBs are in star-forming regions.
Incase, we assume that stars in the early universe are at extragalactic
 distances then GRB events should be correlated with history of the star
formation. This may be true upto $z<2.5$ but it is not true
for $z>2.5$ (Ruiz, {\it et al.}, 2002). Hence, it is more likely that GRB events might have occurred in the back ground of 
the host galaxy which has star-forming region. It is also not
ruled out that it may be mere coincidence with the GRB events
that might have occurred in the star-forming regions of the
host galaxy. Right now, we can not answer the last question, because
aim of the present study is to reveal the source and explain the physics 
behind the central engine of the GRB. 
 
It is known from solar flare observations that before eruption
of the flare, gradients of magnetic fields occur over the solar
surface. It would be interesting to get observational information regarding
development of any such strong gradients of magnetic fields (inferred from the Faraday polarization data in 
radio domain), at the site of GRB flare, before it's eruption.
In the present study we phenomenologically modeled GRB as a primordial
flare phenomenon. However, detailed solution of MHD equations in the 
environment of the early universe is essential 
in order to understand the GRB phenomenon completely.  

{\em The most important finding from the present study is that the 
energy-redshift relation obtained from the simple flare mechanism is similar 
to the observed energy-redshift relation} which will help in finding the 
unknown redshifts of other observed GRB events.

In summary, this study indicates that solar like transient MHD phenomena,
especially primordial flares in the inter galactic or inter galaxy cluster
region, may be most promising energy source for central engine
which triggers the GRB events. Though most of the observed properties of GRB phenomena
and solar flare phenomena are similar, additional observational informations 
such as signature of the gradients in the extragalactic magnetic fields at
the site of production of GRB is required 
in order to prove our proposed conjectures in this study.

\begin{center}
{\Large\bf Acknowledgements}
\end{center}
\noindent This paper is dedicated to my beloved parents who constantly encouraged my
research carrier when they were alive.

\begin{center}
{\Large\bf References }
\end{center}

\noindent Antonelli, et al. 2000, ApJ, 545, L39

\noindent Bagchi {\em et al}, J. 1998, MNRAS, 296, L23

\noindent Band, D.L, Hartmann, D. H and Shaefer, B.E, A\&A, Supp. Ser, 138, 481, 1999

\noindent Bloom, J. S., Kulkani, S. R \& Djorgovski, S. G. 2002, Astron.Jour, 123, 1111

\noindent Brown, M. R. 1999, Phys. Plasma., 6, No 5, 1717 

\noindent Fenimore, E. E., \& Ramirez-Ruiz, E. 2000, astro-ph0004176

\noindent Giovanelli, R. 1998, Astron.Jour, 116, 2632

\noindent Haisch, B., \& Strong, K. T. 1991, Adv. Space. Res. Vol 6, No 8, pp 47-50

\noindent Hood, A. W. 1986, {\em Solar System Magnetic Fields}, edt.,

\noindent E.R. Priest, (D. Reidel Publishing Company), p. 80

\noindent Kroneberg, P. P. 1994, Rep. Prog. Phys., 57, 325 

\noindent Larson, B, B. 1999, (ESA, SP145 ), p. 13

\noindent Lenters, G. T., \& Miller, J. A. 1998, ApJ, 493, 451

\noindent Lesch, H., \& Chiba, M. 1997, Fundamentals of Cosmic Physics, 18, No 4, 278 

\noindent McLaughlin, G. C., Wijers, R. A. M. J., Brown, G. E. \&  Bethe, H. A. 2002, ApJ, 567, 454

\noindent Mirabal, {\em et.al,} 2002, ApJ, 578, 818

\noindent Parker, E. N. 1994, {\em Spontaneous Current Sheets in Magnetic Fields}, (Oxford Univ Press ), p. 286

\noindent Piran, T. 1999, Phys. Rep, 314, 575

\noindent Piro, L., et al. 2000, Science, 290, 955

\noindent Press, W, H., Teukolsky, S. A., Vellerling, W \& Flannery, B. 1992, in Numerical recipes in C, Camridge Univ Press, p. 664

\noindent Priest, E. R. 1981, {\em Solar Flare Magnetohydrodynamics}, edt., E. R. Priest, ( Gordon \& Breach Science Publishers ), p. 14

\noindent Qu Qin-Ye \&  Wang Zhan-Ru. 1977,  Chinese Astronomy, 1, 97

\noindent Schaefer, B. E., ApJ, 511, L79, 1999

\noindent Stecker, F. W., \& Frost, K. J. 1973, Nature, Physical scienices, 245, 70

\noindent Taylor, G. B., \& Perley, R. A. 1993, ApJ, 416, 554

\noindent Tsuneta, S., \& Naito, T. 1998, ApJ, 495, L67

\noindent Vahia, M., \& Rao, A. 1998, A\&A, 207, 55

\noindent Vallee, J. P., {\em et al}, 1987, ApJL, 25, 181

\noindent Ruiz, {\it et al.}, 2002, in "Cosmology and Particle Physics", p. 457

\noindent Woosley, S. E. \& Weaver, T. A. 1995, ApJs, 101, 181

\noindent Zirin, H. 1988, in {\em Astrophysics of the Sun}, p. 378

\clearpage
\begin{figure}[h]
\begin{center}
\psfig{figure=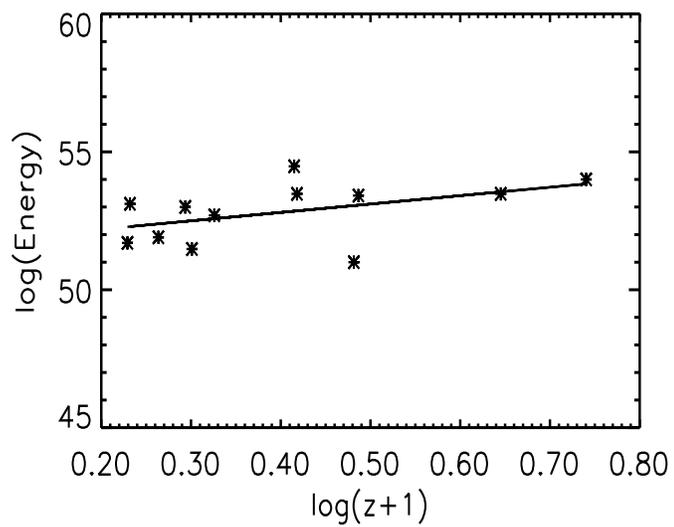,height=8.0cm,width=10.0cm}
\end{center}
\caption{The cosmic GRB  energy versus redshift values in
the logarithmic scale. The observed values are denoted by signs
of star and continuous line is a linear least square fit of the
form $Y = C + MX$, where $Y =log(energy)$, $X=log(z+1)$ and $z$ is
the redshift, to the observed data.}
\end{figure}


\begin{figure}[h]
\epsfig{figure=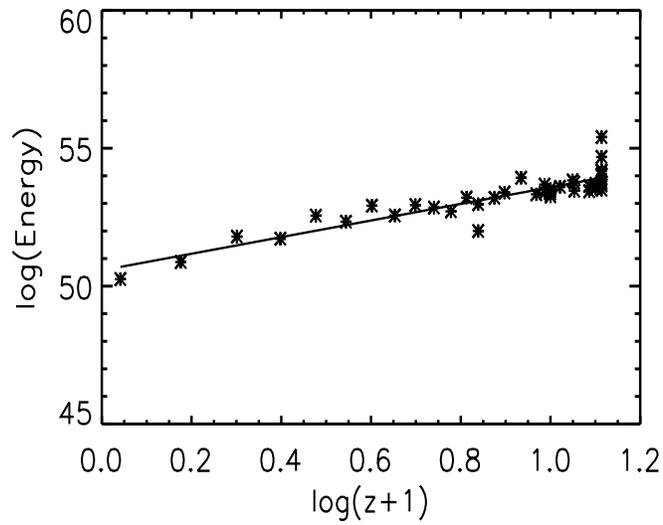,height=8.0cm,width=10.0cm}
\hspace{1cm}
\caption{The cosmic GRB  energy versus redshift values in 
the logarithmic scale. The inferred values (taken from
Fennimore and Ramirez-Ruiz, 2000)are denoted by signs 
of star and continuous line is a linear least square fit of the 
form $Y = C + MX$, where $Y =log(energy)$, $X=log(z+1)$ and $z$ is 
the redshift, to the observed data. \label{fig2}}
\end{figure}

\end{document}